\documentclass[aps,prb,twocolumn,superscriptaddress,showpacs]{revtex4}
\usepackage{graphicx}

\begin{document}
\title{Picosecond Nonlinear Relaxation of Photoinjected Carriers \\
in a Single GaAs/Al$_{0.3}$Ga$_{0.7}$As Quantum Dot}

\author{T.~Kuroda}
\affiliation{Department of Physics, Tokyo Institute of Technology,
Meguro-ku, Tokyo 152--8551, Japan}
\author{S.~Sanguinetti}
\affiliation{I.N.F.M. and Dip. di Scienza dei Materiali, Universit\'{a} di
Milano Bicocca, Via Cozzi 53, 20125 
Milano, Italy} 
\affiliation{Nanomaterials Laboratory, National Institute for
Material Science, 1-2-1 Sengen,
Tsukuba 305-0047, Japan}
\author{M.~Gurioli}
\affiliation{I.N.F.M. and Dip. di Scienza dei Materiali, Universit\'{a} di
Milano Bicocca, Via Cozzi 53, 20125
Milano, Italy}
\author{K.~Watanabe}
\affiliation{Nanomaterials Laboratory, National Institute for 
Material Science, 1-2-1 Sengen, Tsukuba 305-0047, Japan}
\author{F.~Minami}
\affiliation{Department of Physics, Tokyo Institute of Technology,
Meguro-ku, Tokyo 152--8551, Japan}
\author{N.~Koguchi}
\affiliation{Nanomaterials Laboratory, National Institute for 
Material Science, 1-2-1 Sengen, Tsukuba 305-0047, Japan}

\date{\today}

\begin{abstract}
Photoemission from a single self--organized GaAs/Al$_{0.3}$Ga$_{0.7}$As
quantum dot (QD) is temporally resolved
with picosecond time resolution. The emission spectra consisting of the
multiexciton structures are observed to
depend on the delay time and the excitation intensity. Quantitative
agreement is found between
the experimental data and the calculation based on a model which characterizes 
the successive relaxation of multiexcitons.
Through the analysis we can determine the carrier relaxation time 
as a function of population of photoinjected carriers. 
Enhancement of the intra--dot carrier relaxation is demonstrated to be due
to the carrier--carrier scattering
inside a single QD.
\end{abstract}


\maketitle

Semiconductor quantum dots (QDs) are designable mesoscopic atoms easily
integrable in bulk electronics. These
zero--dimensional structures, being solid state systems with an atomic--like
density of states, have attracted
large interest as possible efficient replacements of other quantum
heterostructures for standard optoelectronic
devices. Recently a more general prospective has been addressed, aiming to
exploit the specific features of QDs
for a new generation of novel devices. Examples of this new trend are in the
field of quantum
cryptography \cite{MRG01} and quantum computing \cite{BHH01}, based on
single QD (SQD) devices. Within this
framework it is obvious that the main effort in QD research has been focused
on the comprehension and the
control of SQD properties. Studies of the phase relaxation \cite{BKM01}, the
spectral diffusion \cite{SWB00} and 
the carrier--carrier correlation \cite{BWS99,BSH00,DGE00,DRG00} 
are nowadays the focus of attention in QD basic research. 

In this paper we study the ultrafast nonlinear properties on the carrier
relaxation inside a single QD. The
SQD emission transients are resolved with ps time resolution and interpreted
in terms of the successive
transition between the correlated few--exciton states.
Quantitative analysis on the SQD emission  
allows us to determine the temporal dependence of the carrier population
confined in a single dot.
This permits us to show a strict correlation between 
the intrinsic carrier energy relaxation 
and the number of photoinjected carriers. We demonstrate that the
intra--dot carrier--carrier
scattering process is responsible for the efficient carrier relaxation in
semiconductor QDs.

The samples are GaAs/Al$_{0.3}$Ga$_{0.7}$As self--assembled QDs grown by
modified droplet epitaxy (MDE)
\cite{KTC91,WKG00}. MDE is a non--conventional growth method for
self--assembling QDs even in lattice matched
systems \cite{WKG00}. By modifying the surface reconstruction and the adatom
mobility it is possible to obtain QD
samples with a density as low as $\approx 6 \times10^{8}$~cm$^{-2}$.
Surface and cross sectional high resolution scanning electron microscope
images demonstrate
the formation of pyramidal shape nanocrystals of 16~nm height and 20~nm base
\cite{WTG01}.

The optical measurement was performed with a fs mode--locked Ti--sapphire
laser of 76--MHz repetition rate.
A second--harmonic beam of the laser output ($\lambda =400$ nm) was used for
excitation.
This beam was loosely focused on
the sample by a conventional lens of $f=$ 200 mm. The emitted signal was
collected by a microscope objective of
N.A. = 0.5. The present configuration allows one to irradiate the sample
homogeneously inside the detection spot,
and to determine the power density with precise accuracy. The spatial
resolution was $\sim 1.2 \mu$m, causing
$\sim 7$ QDs to be collected on average. The signal was dispersed by a
polychromator, and detected by a synchronously
scanning streak camera. The temporal and spectral resolution was 15~ps and
0.8~meV, respectively. The sample was
attached to a cold finger of a He--flow cryostat. All experiments were
performed at 3.5~K.

\begin{figure}
\includegraphics[width=8cm,keepaspectratio]{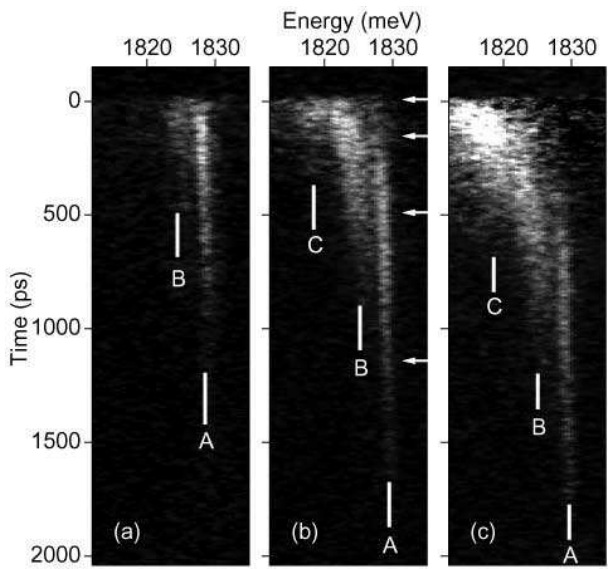}
\caption{\label{F1}Temporally-- and spectrally resolved images of a SQD
emission at various excitation
power densities:
P$_{exc}=47$~nJ/cm$^2$ (panel a), 150~nJ/cm$^2$ (panel b), and 470~nJ/cm$^2$
(panel c).
Three emission components characteristic to the SQD spectra are denoted 
by A, B, and C.
}
\end{figure}

Figure \ref{F1} shows a series of time--resolved SQD emission in a linear
gray scale at various excitation
powers, P$_{exc}$. These emissions 
originate from the recombination of an electron (and a hole) which 
is in the lowest--energy QD state. 
As is clearly shown, a highly nonlinear behavior and
peculiar temporal dynamics appear on the SQD emission.
At low excitation, P$_{exc}=47$~nJ/cm$^{2}$, Fig.~\ref{F1}(a) exhibits a
narrow line,
defined as A, and a weak band, defined as B, at the low--energy side of A.
Both components decrease monotonically with time, although the B band decays
much faster than the A line.
With increasing excitation power to 150 nJ/cm$^{2}$ (Fig.~\ref{F1}(b)), the
B band appears as strong as A,
and an additional rapid--decay component, defined as C, emerges at the lower
energy.
Simultaneously, the temporal shape of A tends to be stretched, and the
finite rise profile is evidently resolved.
After nearly 500 ps the spectral evolution is almost identical with that of
the lower excitation case. On
increasing the irradiation, as in Fig.~\ref{F1}(c), the broad C band
dominates the transient stage of emission,
and the rise of A and B is found to be slower. It is worth noting that the
present features
are common to all the SQD emissions, independent of their energy or dot size.

\begin{figure}
\includegraphics[width=8.6cm,keepaspectratio]{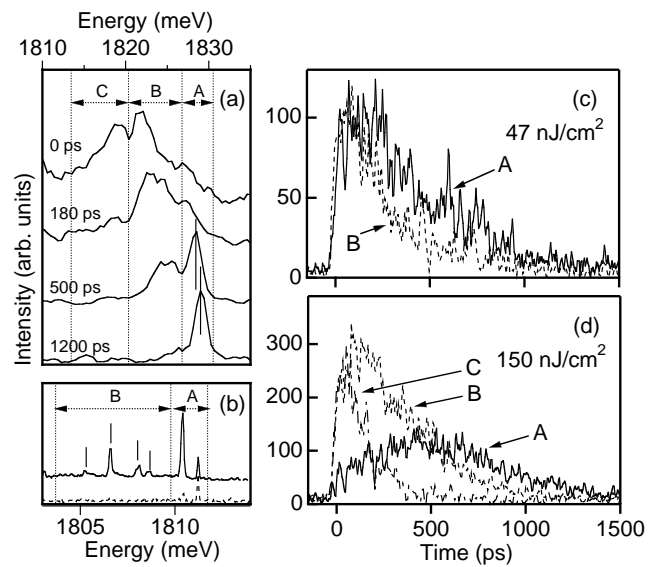}
\caption{\label{F2} Panel a: transient SQD emission spectra 
at P$_{exc}=150$ nJ/cm$^2$
for various delay times, which are presented by the white arrows in
Fig.~\ref{F1}(b).
Panel b: high--resolution SQD spectra with cw excitation
at 2~kW/cm$^2$ (solid line) and 70~W/cm$^2$ (dotted line). Panels c and d:
temporal evolution of the SQD emission
bands of A, B, and C, at P$_{exc}=47$~nJ/cm$^2$ and 150~nJ/cm$^2$,
respectively. Spectral windows for the signal
integration are shown in (a). }
\end{figure}

The temporal behavior of SQD emission is characterized by the transient
PL spectra reported in Fig.~\ref{F2}(a).
We selected four spectra from the PL trace at P$_{exc}=150$ nJ/cm$^{2}$,
indicated by the white arrows
in Fig.~\ref{F1}(b). The spectra show that: (i) the broad C band together
with the B band dominates
the early stage of the emission;
(ii) the C band rapidly fades away with time; and finally, (iii) in the long
time limit, the B band is
totally replaced by the A line. Temporal development of these emission bands
are reported in Figs.~\ref{F2}(c)
and \ref{F2}(d) for two different excitation powers. In the transient
spectra, we also resolve a red shift of the
A line ($\sim$0.8 meV, presented by vertical lines in Fig.~\ref{F2}(a)) and
a narrowing of the B band with
increasing time. These observations suggest that A and B originate from
multiple components which are not
spectrally resolved. This speculation is supported by the high--resolution
SQD spectra
with cw excitation, reported in Fig. \ref{F2}(b). In fact, the
high--resolution spectra exhibit two
lines split by 0.8~meV in the A--line region,
while four lines characterize the B--band region. When the irradiation is
weak enough,
the highest--energy line appears alone, and with increasing irradiation,
other lower--energy lines take place.
Note that a very similar spectral profile is obtained for Figs.~\ref{F2}(a)
and \ref{F2}(b),
although a different QD was captured.

The transient emissions are therefore characterized by three components: 
the long
lasting, high energy A line at 1828 meV, the B line at 1823 meV, and the
fast and broad C band whose energy
depends on the excitation power and delay time. This peculiar evolution can
be attributed to multiexcitonic
effects. The presence of more than one exciton in the QD determines the
splitting and the red shift of the
emission energy of the QD ground state transition. In fact, the carrier
exchange interaction leads to a
progressive decrease of emission energy when increasing the number of
excitons in the QD, although a certain
reduction is expected due to correlation effects \cite{WFZ01}. Thus, when
the number of excitons
increases, more lines appear in the low energy side of the fundamental
single exciton transition. 
Within this framework we
attribute the doubly degenerate A line to single exciton and biexciton
recombinations, and the four--fold
degenerate B band to the recombinations when the total number of excitons is
between 3 and 6. A larger number of
excitons captured in the QD give rise to the C band. These attributions
reproduce the power dependence of the
SQD emissions well, as will be discussed later. 

It is noteworthy that the multiexciton complex has been first explored
in bulk indirect semiconductors weakly doped by impurities 
\cite{Sauer73,ThewaRost78}.
In this case multiple carriers are bound by the Coulomb interaction,
and their eigen-state is described by the \textit{shell} model,
which specifies the relative motion of carriers \cite{Kirczenow77}.
For the case of QDs, on the other hand, 
multiple carriers are bound mainly due to the quantum confinement, 
and the \textit{shell} structure is simply given by the confinement sublevels
determined for a single carrier.
The carrier--carrier correlation reflects in the renormalization
of the single carrier levels.
In the present QDs, the energy split between
the lowest energy shell (\textit{s}~shell) 
and the second-lowest one (\textit{p}~shell)
is $\sim$ 80 meV,
and we focus on the recombination of an electron (and a hole) 
occupied in the \textit{s}~shell, which is modified by the correlation
with the other carriers.

To perform a quantitative discussion, we analyze the time evolution of the
SQD PL using a model based on
the successive transitions between the multiexciton states. Since our
observations suggest that
the transient SQD spectrum is solely determined by the population of
excitons at a specific time,
the emission
dynamics are described by a set of rate equations \cite{Sauer73,DRG00},
$d\rho_{i}/{dt}=-\Gamma_{i}\rho_{i}+\Gamma_{i+1}\rho_{i+1}$,
where $\rho_{i}(t)$ is the probability to find $i$~excitons inside the QD,
and $\Gamma_{i}$ is the $i$X
transition rate -- hereafter $i$X indicates the recombination with
$i$~excitons inside the QD. We
also adopt the realistic assumption that the number of photoinjected
carriers are statistically distributed,
and it is given by the Poisson distribution, so that
$\rho_{i}(t=0)=e^{-\bar{\rho}}\bar{\rho}^{i}/i\text{!}$,
where $\bar{\rho}$ represents the average number of photoinjected carriers
\cite{Sauer73}.

\begin{figure}
\includegraphics[width=8.6cm,keepaspectratio]{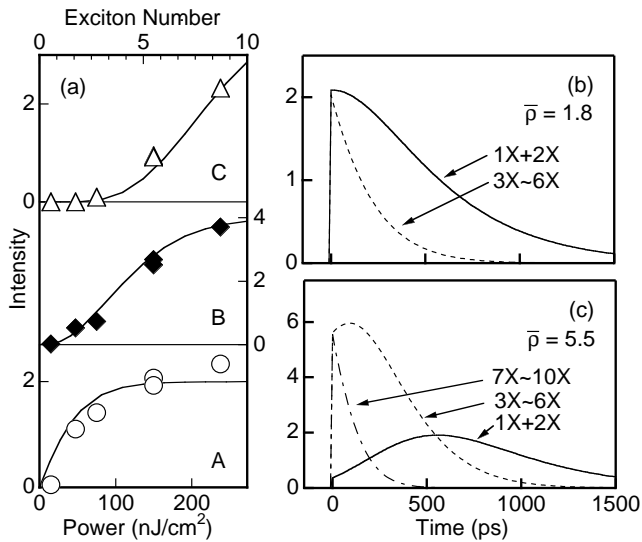}
\caption{\label{F3} Panel a: excitation power dependence of time--integrated
intensities of the A-- (circles),
B-- (diamonds), and C band (triangles). Solid lines indicate the fit with
$\xi =1.85 \times 10^{-11}$
cm$^{2}$. Panels b and c: Model calculations for the temporal development of
SQD emission.
These curves are compared with the experiment, Figs.~\ref{F2}(c) and \ref{F2}(d).
The relation between the magnitudes of $\bar{\rho}$ for panel b and panel c 
corresponds to that of the excitation powers in the experiment.
The ratio of the vertical scales is identical with that in Fig.~\ref{F2}. }
\end{figure}

The efficiency of photoinjection was extracted by analyzing the
time--integrated (TI) intensities
of the SQD emissions.
Variations of the TI intensities for the A--, B--, and C emission bands are
presented in
Fig.~\ref{F3}(a), together with the calculations based on the above model.
The best fit to the data gives
an efficient carrier--capturing cross section, $\xi$, of $1.85 \times
10^{-11}$ cm$^{2}$.
With the use of $\xi$,
the number of photoinjected carriers is given by the multiplication of $\xi$
with the photon flux,
or the carrier density at the sample surface.
The present procedure allows us to determine the carrier population inside
the QD.
It was found that the magnitude of $\xi$ is consistent with a result of
the PL yield measurement of the present sample, and of the same order with
that of the InAs/GaAs QDs
reported in Ref.~\onlinecite{DRG00}.

To interpret the temporal evolution of the SQD emission, $\rho_{i}(t)$ is
solved numerically.
For simplicity we assume that the transition rate of the multiexciton state
is given by
the summation of the relevant
single--carrier recombination rates. In this treatment, dynamics of up to 10
excitons are specified using the
three lowest inter--shell (\textit{s}--, \textit{p}--, and another higher
shell) transition rates as free
parameters. In Figs.~\ref{F3}(b) and \ref{F3}(c), the best fits of the data
(Figs.~\ref{F2}(c) and \ref{F2}(d))
are presented, 
where the transition times for the \textit{s}--, \textit{p}--, and the
higher energy shell are derived as
400~ps, 600~ps, and 300~ps, respectively, by an accuracy of 20 \%. Excellent
agreement between calculation
and the data supports the validity of the assumption that the multiexciton
transition strength, \textit{i.e.},
overlapping between the few electron--hole pairs, is not strongly modified
by the carrier--carrier correlation.

\begin{figure}
\includegraphics[width=6.5cm,keepaspectratio]{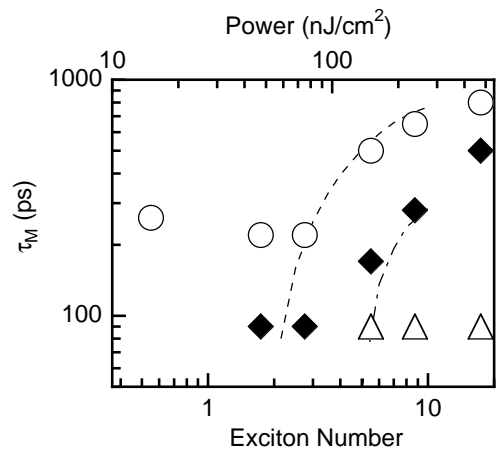}
\caption{\label{F4}
Variation of a phenomenological rise time, $\tau_M$,
as a function of average number of photoinjected carriers (or excitation
power density) for the A-- (circles),
B-- (diamonds), and C band (triangles). 
Model results of $\tau_M$ for the A (B) band are indicated by a broken 
(broken dotted) line.
}
\end{figure}

The rise characteristics of the SQD emissions are shown in Fig.~\ref{F4}.
Here we plot a phenomenological rise time, $\tau_M$, by measuring the
time delay needed
for each emission band to reach its maximum, as a function of average number
of photoinjected carriers.
Note that $\tau_M$ is roughly three to four times larger than the
conventional rise time, $\tau_R$, obtained by a single--exponential fit.
The C band indicates the fastest rise time ($\tau_{M}=85 \pm10$~ps), close
to our time
resolution limit. The A band and the B band show an increase in rise time
when the exciton number is larger than two. This delay feature is
understood by the present model
based on the successive relaxation of multiexcitons: the emission of the $i$X
band starts
only when the internal excitons decrease to $i$,
through several recombinations after pulse excitation, thus resulting in a
delay of the $i$X onset time.
The observed behavior of the emission rise for $\bar{\rho}>2$ is well
reproduced by the calculation
using the parameters described above.
On the other hand, in the case of low carrier injection,  $\bar{\rho} < 2$,
the experimental rise time is instead longer than the predicted one.
It is therefore suggested that
the intra--dot carrier relaxation to the ground--state configuration is
evidently slowed down
in the absence of multiple carriers inside the QD.
This phenomenology is strictly related to the well--known PL characteristics
observed by the ensemble QD experiment.
In this case,
the emission rise time, $\tau_R$, 
increases to several 10 ps at weak excitation density. 
With increasing density, on the other hand, the PL rise is observed to be fast. 
The commonly accepted interpretation is that 
the intra--dot relaxation is enhanced due to the
Auger--like scattering of
photoinjected carriers involving continuum electronic states 
\cite{MPF99,UMA98,RHF00}.

Our SQD data is consistent with the result of ensemble QD measurements,
which generally show
a monotonic decrease in PL rise time with photoinjection intensity 
\cite{SWT02}.
In addition, by resolving the SQD emission dynamics 
we obtain a different and  more precise picture of the intrinsic relaxation 
process.
The speeding up of
the carrier energy relaxation occurs whenever the number of excitons 
in the SQD is larger than few unities.
At low carrier injection, when
there is less than two excitons per QD 
(in this case the B band is weak or not resolved),
the carrier relaxation
to the ground state is relatively slow and almost independent of the
excitation power.
The B band already shows a fast rise at the minimum injection needed for
detection.
With increasing photoinjection, the C band becomes the dominant contribution
of the SQD emission
(the A and B bands appear only at very long delay times).
Contribution of the all SQD emission bands leads to the monotonic decrease
in the PL rise time, found in the ensemble averaged measurement.
These results demonstrate
that the shortening of the rise time of the QD emission stems from a
multiexciton recombination,
which cannot be resolved in standard PL of the QD ensemble, due to the large
inhomogeneous broadening.
At the same time our data
show that the Auger--like processes effective in the QD relaxation should be
associated to carrier--carrier
scattering inside the QD \cite{FerBas99,SWT02}, rather than that involving the
carrier population in the wetting layer
and/or barrier states, as usually considered in the literature
\cite{MPF99,UMA98,RHF00}.

In conclusion, we have determined the nonlinear carrier dynamics in a single
quantum dot by means of PL
measurements with picosecond time resolution. Effects of the multiexciton
dynamics leads to strong nonlinearities
in the SQD emission. The observed behavior stems from successive relaxation
dynamics of multiexcitons. Moreover,
the efficient intra--dot relaxation has been shown to be due to the
nonlinear interaction between carriers inside
a QD. We believe that the investigation of transient SQD emissions allows us
to get deeper understandings on the
ultrafast carrier dynamics in zero dimensional semiconductor systems.


\end{document}